\documentstyle[12pt]{article} \begin{document}
\overfullrule 0 mm
\language 0
\centerline {{ \bf THE LORENTZ-DIRAC EQUATION:}}
\centerline {{ \bf ONE MORE PARADOX OF PREACCELERATION}}
\vskip 0.3 cm

\centerline {{  Clexander A. Vlasov}}
\centerline {{  High Energy and Quantum Theory}}
\centerline {{  Department of Physics}}
\centerline {{  Moscow State University}}
\centerline {{  Moscow, 119899}}
\centerline {{  Russia}}
\vskip 0.3 cm

{\it One more paradox of classical Lorentz-Dirac
preaccelerative  solution is found: the formation of the event
horizon }

03.50.De
 \vskip 0.3 cm

Since the famous Dirac's paper on relativistic radiation reaction in
classical electrodynamics, many textbooks and research articles were
published on that theme. Among them are [1-8], where one can find 
the discussion of the related problems: mass renormalization, 
runaway solutions and the use of the advanced interaction.

There is the opinion in the literature  that the method of backward
integration in time together with preacceleration can solve some
problems of Lorentz-Dirac equation.

However the use of preaccelerative solutions produces a large amount
of paradoxes (some of them are discussed in [7,8]).

In our paper we present one more paradox of preacceleration: the
formation of unavailable area of initial data, i.e. the formation of
"event" horizon, absent in classical equation without radiation
reaction.

Consider for simplicity the nonrelativistic case (with units $c=1$).

Let the point particle with mass $m$ and charge $e$ move (following
the classical nonrelativistic equation without radiation reaction)
under the influence of the external force $F$ along the $x$-axis:
$$ {d^2 x \over dt^2} ={F(x)\over m} \eqno(1)$$
Let's take the force $F(x)$ in the form of a positive  step:
 $$ F =\left
  (\matrix{0, &-x_{0}<x \cr mA &-x_{0}<x<0\cr 0,
&0<x\cr}\right ) \eqno(2)$$
here $A>0,\ \ \ x_{0}>0$.
In the method of backward integration one must take the "final" data
(i.e. for $t\to \infty$) and zero final acceleration. Thus for (2) we
assume free motion of the particle in the future with velocity $v$:
$$x=vt \eqno(3)$$

Suppose that the point $x=0$ is achieved at $t=0$, and the point
   $x=-x_{0}$ - at  $t=t_{0}<0$ (the value of $t_{0}$ we shall find
below).

Then the integration of (1-2,3) with appropriate boundary conditions
 (for the second order equation (1) the position and the velocity of
  particle must be continuous) yields:
   $$ x =\left (\matrix{vt, &t>0 \cr vt+At^2/2
&t_{0}<t<0\cr ut+b, &-\infty<t<t_{0}\cr}\right ) \eqno(4)$$

Values of $u,\ b,\ t_{0}$ are determined from the
matching conditions: $$u= \sqrt{v^2-2x_{0}A},\ \ t_{0}={-v+u \over
A},\ \ b=-x_{0}-ut_{0}\eqno(5)$$
Following (4) in the regions free of force ($x>0$ and $x<x_{0}$ 
) the particle motion is free.

Consider the   case of small velocity ${dx \over dt}=u$ at
the point $x=-x_{0}$:  $$u=\sqrt{v^2-2x_{0}A}=\epsilon,\ \ \  
\epsilon \to
 0\eqno(6)$$
We see that though the value of $u$ is small, the
particle can reach from the future all points on the $x$-axis:
$-\infty<x<+\infty$.

Consider now the same particle with the same final data moving under
 the same external force $F$ (2) but obeying the nonrelativistic
Lorentz-Dirac equation
 $${d^2 x \over dt^2}-k{d^3 x \over dt^3}
={F(x)\over m} \eqno(7)$$ here $k$ ($k\approx
x_{cl}$-the classical radius of a particle, $k>0$) and the second
term on L.H.S. of (7) deal with the radiation force.

As the equation (7) is of the third order, the acceleration of
the particle also must be continuous. Then the solution of the above
problem for equation (7) with appropriate boundary conditions yields
(the point $x=0$ is achieved at $t=0$, and the point
   $x=-x_{0}$ - at  $t=t_{1}<0$ )

  $$ x =\left
  (\matrix{vt, &t>0 \cr (v+kA)t+At^2/2
+k^2A(1-\exp{(t/k)})&t_{1}<t<0\cr ut+b+c\exp{(t/k)},
&t<t_{1}\cr}\right ) \eqno(8)$$

Values of $u,\ b,\ c,\ t_{1}$ are determined from the
matching conditions:
  $$-x_{0}=ut_{1}
+b+c\exp{(t_{1}/k)}\eqno(9a)$$
$$-x_{0}=(v+kA)t_{1}+A(t_{1})^2/2
+k^2A(1-\exp{(t_{1}/k)})\eqno(9b)$$
$$v+kA+At_{1} -kA\exp{(t_{1}/k)})=
u+(c/k)\exp{(t_{1}/k)}\eqno(9c)$$
 $$A-A\exp{(t_{1}/k)}
=(c/k^2)\exp{(t_{1}/k)}\eqno(9d)$$
Term with exponent in (8) for $x<-x_{0}$ (i.e. "before" the action of
the force) describes the effect of preacceleration.

From equation (9d) immediately follows that $c>0$ for $t_{1}<0$.

Consider the   case of zero  particle velocity ${dx \over
dt}$ at the point $t=t_{1}<0,\ \ x=-x_{0}$:
 $${dx \over dt}=u+(c/k)\exp{(t_{1}/k)}=0 \eqno(10)$$
In (10)   the
value of $u$ must be negative because the value of   $c$ is
positive.

Equation (10) with the help of the system (9) can be rewritten as
  $$z^2/2-1-p= \exp{(z)} (z-1)
\eqno(11)$$ here $z=t_{1}/k<0,\ \ p=x_0/(k^2A)>0$.
For our goal it is important to note that the equation (11) always
has  solution for $z=z^{*}<0$ and this solution varies weakly with
small changes in parameters of the problem under consideration.

Consequently if we consider the similar to (6)  case
of small particle velocity  at the point $x=-x_{0}$:  $${dx \over
dt}(t_2)=\epsilon,\ \ \ \epsilon \to 0\eqno(12)$$
then $t_2 \approx t_1$ and, as for (10),
  $$u<0,\ \ \ c>0 $$.  These unequalities lead to the conclusion that
the particle velocity in the free of force region ($x<-x_{0}$) -
${dx \over dt}=u+(c/k)\exp{(t/k)}$, with positive value
$\epsilon$ at $x=-x_{0}$, must inevitably take (with time decrease)
zero value at some moment $t=t_{min}<t_2$ and some point
$x=x_{min}<-x_{0}$.

So all $x$ on the left of $x_{min}$: $x<x_{min}$, become unavailable
on contrary to  solution (6) of the  equation (1),
where all $x$-axis  is available for the particle. Thus the event
horizon is formed: not all initial data, physically possible from the
point of view (1), can be achieved by this method.

 This paradox (as all the other) makes unrealizable the desire to 
solve problems of Lorentz-Dirac equation using preacceleration and 
backward in time integration.

If one takes this point of view than two possibilities follow:

(i) to state that the Lorentz-Dirac attempt to construct relativistic
radiation force is invalid and to search for new equations (see, for
ex., [9]);

(ii) to reject preacceleration and to consider runaway solutions as
real physical solutions of the bound system particle+field.

Perhaps the recent investigations [10,11] of the  radiation
problems for the classical relativistic scalar field speak in favour 
of (ii).

    \vskip 3 mm
  \begin{enumerate}
  \item
  F.Rohrlich, {\it Classical Charged Particles}, Addison-Wesley,
  Reading, Mass., 1965.
\item
A.Sokolov, I.Ternov, {\it Syncrotron Radiation}, Pergamon Press,
    NY,1968. A.Sokolov, I.Ternov, {\it Relativistic
Electron} (in russian), Nauka, Moscow, 1983.
\item D.Ivanenko, A.Sokolov,  {\it Classical Field Theory} (in
russian), GITTL, Moscow, 1949
 \item S.Parrott, {\it
Relativistic Electrodynamics and Differential Geometry},
Springer-Verlag, 1987.

 \item C.Teitelboim, Phys.Rev.,1970, D1, p.1572; D2, p.1763.
\item  E.N.Glass, J.Huschilt and G.Szamosi, Am.J.Phys., 1984, v.52,
p.445.
\item S.Parrott, Found.Phys., 1993, v.23, p.1093.
\item W.Troost et al., Internet: hep preprints database, preprint
hep-th/9602066, 1996

  \item W.B.Bonnor, Proc.Roy.Soc. (London). Ser. A, 1974, v.337,
p.591; A.Lozada et al, J.of Math. Phys., 1983, v.30, p.1713.

\item D.Vollick, Phys.Rev., 1995, D52, n.6, p.3576
\item A.A.Vlasov, Teor.Mat.Fiz., 1996, 109, n.3, p.464.

\end{enumerate}

 \end{document}